\documentclass[12pt,preprint]{aastex}

\def\mf{m_f}
\def\nf{n_f}
\def\ff{f_f}
\def\sf{\sigma_f}
\def\rf{\rho_f}
\def\vf{v_f}

\def\pf{p_f}
\def\m12{M_{12}}
\def\msun{M_{\odot}}

\def\dv{\langle\Delta\vartheta^2\rangle}
\def\beq{\begin{equation}}
\def\eeq{\end{equation}}
\def\fun#1#2{\lower3.6pt\vbox{\baselineskip0pt\lineskip.9pt
  \ialign{$\mathsurround=0pt#1\hfil##\hfil$\crcr#2\crcr\sim\crcr}}}

\def\gap{\mathrel{\mathpalette\fun >}}

\begin{document}

\title{Rotational Brownian Motion of a Massive Binary}
\author{David Merritt}
\affil{Department of Physics and Astronomy, Rutgers University}

\begin{abstract}

The orientation of a massive binary undergoes a random walk due
to gravitational encounters with field stars.
The rotational diffusion coefficient for a circular-orbit binary 
is derived via scattering experiments. 
The binary is shown to reorient itself by an angle of
order $\sqrt{m/M}$ during the time that its semi-major
axis shrinks appreciably, where $M$ is the binary mass
and $m$ the perturber mass.
Implications for the orientations of rotating black holes 
are discussed.

\end{abstract}

\section{Introduction}

Two sorts of Brownian motion are defined by statistical physicists.
Translational Brownian motion is the irregular motion exhibited by a massive
particle as it collides with molecules in a fluid.
Collisions also affect the spin and orientation of particles; 
theories of rotational Brownian motion are concerned with the evolution 
of these quantities due to encounters.
Rotational Brownian motion was first discussed by \citet{deb13,deb29} 
in the context of dielectric theory.
The polarization of a dielectric material is a competition between 
torques due to the imposed field, which tend to align polar molecules,
and collisions, which tend to destroy the alignment.

The likely existence of black holes with masses much greater than
those of stars has led to a renewed interest in Brownian motion among 
astrophysicists \citep{mer01,mim01,dhm01}.
Translational Brownian motion of black holes in galactic nuclei
may be directly observable (e.g. \citet{rei99,bas99}),
allowing constraints to be placed on black hole masses.
Translational Brownian motion may also affect the rate of tidal
disruption of stars by a nuclear black hole (e.g. \citet{you77})
or the efficiency with which a binary black hole 
can interact with stars as it wanders through a nucleus \citep{quh97,mim01}.
In Paper I \citep{mer01}, the translational Brownian motion 
of a massive binary was discussed; it was shown that the amplitude
of the wandering can be increased by a modest factor 
compared with that of a single point mass due to
exchange of energy  between field stars and the binary.

Here we consider the second kind of Brownian motion experienced
by a binary,
the irregular variation of the orientation of the binary's spin axis
due to encounters.
A single field star that passes within a distance $\sim a$ of
the binary, with $a$ the binary's semi-major axis,
will exchange orbital angular momentum with the binary, leading both
to a change in the binary's orbital eccentricity
as well as a change in the orientation of the binary's spin axis.
Repeated encounters will cause the binary's orbital eccentricity
to evolve (usually in the direction of increasing eccentricity) 
and will also cause the orientation of the binary to
undergo a random walk.
The first process has been discussed extensively (e.g. \citet{miv92,qui96});
the second, discussed here for the first time,
will be called rotational Brownian motion,
by analogy with the similar process that occurs in a fluid of polar molecules
(e.g. \citet{mcc80}).

The orbital orientation of a black-hole binary influences the direction of the
spin axis of the single black hole that forms after 
coalescence of the binary due to emission of gravitational radiation
(e.g. \citet{flh98}).
The spin axis, in turn, is thought to determine the plane of the inner
accretion disk which forms around the black hole \citep{bap75}, 
the direction of jets launched from the accretion disk \citep{ree78}, 
etc.
Thus changes in the orientation of a black-hole binary as it coalesces
may have observable consequences.

In \S 2 the conservation equations for interaction of a field star
with a massive binary are presented; most of these have been seen before
and were used in earlier studies to derive the rates of change of the 
binary's separation and eccentricity.
The rotational diffusion equation is presented in \S3 and the order-of-magnitude of the rotational diffusion coefficient $\dv$ is derived.
\S 4 describes the numerical derivation of $\dv$ from scattering experiments.
$\S5$ presents solutions of the diffusion equation and derives a relation
between the degree of hardening of a binary and the change in its orientation.

\section{Encounter Kinematics}

Consider an encounter of a single particle (the ``field star'')
with a binary.
Assume that the binary remains bound during the encounter,
appropriate if the mass of the field star is much less than the
mass of each of the components of the binary
or if the binary is sufficiently hard.
At early times, the field star has velocity ${\bf v}_0$ with
respect to the center of mass of the field star-binary system
and its impact parameter is $p$.
A long time after the encounter, the velocity ${\bf v}$ of the field
star attains a constant value.
Conservation of linear momentum implies that the change 
$\delta{\bf V}$ in the velocity of the binary's center of mass
is given by
\beq
\delta{\bf V} = -{\mf\over\m12} \delta{\bf v}
\eeq
where $\mf$ is the mass of the field star and $\m12=M_1+M_2$
is the mass of the binary.
This relation was used in Paper I in conjunction with scattering
experiments to investigate the translational Brownian motion of a binary due to
encounters.

The energy of the field star-binary system, expressed in terms of
pre-encounter quantities, is
\begin{mathletters}
\begin{eqnarray}
E_0&=&{1\over 2}\mf v_0^2 + {1\over 2}\m12 V_0^2 - {GM_1M_2\over 2a_0}\\
&=& {1\over 2}\mf \left(1+{\mf\over\m12}\right)v_0^2 - {GM_1M_2\over 2a_0}
\end{eqnarray}
\end{mathletters}
with ${\bf V}_0=-(\mf/\m12){\bf v}_0$ 
the initial velocity of the binary's center of mass
and $a_0$ the binary's initial semi-major axis.
After the encounter,
\beq
E= {1\over 2}\mf\left(1+{\mf\over\m12}\right)v^2 - {GM_1M_2\over 2a}
\eeq
and $E=E_0$, so that
\begin{mathletters}
\begin{eqnarray}
\delta\left({1\over a}\right)&=&
{\mf(v^2-v_0^2)\over GM_1M_2}\left(1+{\mf\over\m12}\right)\\
&\approx& {\mf(v^2-v_0^2)\over GM_1M_2}
\label{eq_harden}
\end{eqnarray}
\end{mathletters}
where the latter relation assumes $\m12\gg\mf$.
This result was used by \citet{hil83,hil92}, \citet{miv92},
and \citet{qui96} to compute the hardening rate,
$(d/dt)(1/a)$, of a massive binary due to encounters.

The angular momentum of the field star-binary system about its center of mass, 
expressed in terms of pre-encounter quantities, is
\beq
{\bf \cal{L}}_0=\mf\left(1+{\mf\over\m12}\right){\bf \ell}_0 + \mu_{12}{\bf \ell}_{b0} 
\eeq
where $\mu_{12}\equiv M_1M_2/\m12, {\bf \ell}_0\equiv p{\bf v}_0$ and ${\bf \ell}_{b0}\equiv{\bf \cal{L}}_{b0}/\mu_{12}$
with ${\bf \cal{L}}_b$ the binary's spin angular momentum.
Conservation of angular momentum during the encounter gives
\begin{mathletters}
\begin{eqnarray}
\delta{\bf \ell}_b &=& 
-{\mf\over\mu_{12}}\left(1+{\mf\over\m12}\right)\delta{\bf \ell} \\
&\approx& -{\mf\over\mu_{12}}\delta{\bf \ell}.
\label{eq_dl}
\end{eqnarray}
\end{mathletters}
Changes in $|{\bf \ell}_b|$ correspond to changes in the 
binary's orbital eccentricity $e$ via the relation $e^2=1-\ell_b^2/G\m12\mu^2 a$
\citep{miv92,qui96}.
Changes in the direction of ${\bf \ell}_b$ correspond to changes in the
orientation of the binary, leading to rotational diffusion.

\section{Rotational Diffusion Equation}

Let $F(\theta,\phi,t)d\Omega$ be the probability that the spin axis
of the binary is oriented within the solid angle $d\Omega$ at time $t$.
We seek an equation describing how $F$ evolves with time due to
encounters of field stars with the binary, if each encounter
is assumed to produce a negligibly small change in $\theta$ and $\phi$.
If we imagine that ${\bf \cal{L}}_b$ is initially oriented parallel to the 
$\theta=0$ axis, and that the
encounters leading to changes in the orientation of the binary 
are isotropic in velocity and direction,
then $F$ will evolve in such a way as to remain a function of
$\theta$ alone, $F=F(\theta,t)$.
These are the same assumptions made by \citet{deb29} in his theory
of the rotational Brownian motion of spherical molecules.
Debye showed that the evolution equation for $F$, in the absence
of an external torque, is
\beq
{\partial F\over\partial t} = {1\over \sin\theta}{\partial\over\partial\theta}\left(\sin\theta{\dv\over 4}{\partial F\over\partial\theta}\right)
\label{eq_evol1}
\eeq
with $\dv$ the rotational diffusion coefficient,
\beq
\dv = \int \Psi(d\Omega,d\Omega')\vartheta^2d\Omega';
\eeq
$\Psi(d\Omega,d\Omega')d\Omega'$ is the probability that, during
a unit interval of time, a binary whose angular momentum ${\bf \cal{L}}_b$ 
is directed toward $d\Omega$ will reorient itself such that ${\bf \cal{L}}_b$ lies
within $d\Omega'$, and $\vartheta$ is the angular separation
between $d\Omega$ and $d\Omega'$.

From its definition, $\dv$ is the sum, over a unit interval of time,
of $(\delta\vartheta)^2$ due to encounters with field stars.
It can be expressed in terms of mean changes in the components of the
binary's angular momentum as follows.
Let the binary be initially oriented with its spin vector along the $z$ axis,
${\bf \ell}_{b0} = \ell_{b0}{\bf e}_z$.
After an encounter, the angle $\vartheta$ between initial and final
spins is
\beq
\cos\vartheta = {\ell_{b,z}\over \ell_b} = {\ell_{b,z}\over\sqrt{\ell_{b,x}^2+\ell_{b,y}^2+\ell_{b,z}^2}}.
\label{eq_cost}
\eeq
Expanding equation (\ref{eq_cost}) to second order in $\vartheta$
and taking averages, we find
\beq
\dv={\langle\Delta \ell_{b,x}^2\rangle + \langle\Delta \ell_{b,y}^2\rangle\over \ell_{b0}^2}
\eeq
which can further be expressed in terms of changes in the 
angular momentum components of the field star 
via equation (\ref{eq_dl}) as
\begin{mathletters}
\begin{eqnarray}
\dv&=& {\mf^2\over\mu_{12}^2} {\langle\Delta \ell_x^2\rangle + 
\langle\Delta \ell_y^2\rangle\over \ell_{b0}^2}\\
&=&{\mf^2\over GM_1M_2\mu_{12}a(1-e^2)}
\left[\langle\Delta \ell_x^2\rangle + \langle\Delta \ell_y^2\rangle\right].
\label{eq_diff1}
\end{eqnarray}
\end{mathletters}
The last relations assume $\m12\gg\mf$.
Specializing to the case of an equal-mass, circular-orbit binary,
equation (\ref{eq_diff1}) becomes
\beq
\dv =
16\left({\mf\over\m12}\right)^2
{\langle\Delta \ell_x^2\rangle + \langle\Delta \ell_y^2\rangle \over
G\m12a}.
\label{eq_diff2}
\eeq
Equation (\ref{eq_diff2}) is a prescription for computing the rotational
diffusion coefficient from scattering experiments.
This prescription will be applied in the next section.

We can estimate the order of magnitude of the rotational diffusion coefficient
by noting that $\sqrt{G\m12/a}=V_{bin}$, the relative velocity of 
$M_1$ and $M_2$,
and that a single close encounter produces a change in $\ell$ of the field star
of order $V_{bin}\times a$.
The rate of close encounters is 
$\sim 2\pi G\m12\nf a/\sf$ with $\nf$ and $\sf$ the number density
and 1d velocity dispersion of the field stars (Paper I).
Thus
$\langle \Delta \ell^2\rangle \approx (aV_{bin})^2\times  2\pi G\m12\nf a/\sf$
and
\beq
\dv \approx 32\pi {\mf\over\m12} {G\rf a\over\sf}
\eeq
where $\rf\equiv \mf\nf$.

The same encounters that induce rotational diffusion will also 
cause the binary to harden, according to equation (\ref{eq_harden}).
The hardening rate may be written
\beq
{d\over dt}\left({1\over a}\right) = {G\rf H\over\sf}
\label{eq_defH}
\eeq
with $H\approx 15$ for a hard binary (e.g. \citet{qui96}).
Thus
\beq
\dv \approx {32\pi\over H}{\mf\over\m12}\left[a{d\over dt} \left({1\over a}\right)\right] \approx {32\pi\over H}{\mf\over\m12}{1\over t_{hard}}
\label{eq_compare}
\eeq
with $t^{-1}_{hard}\equiv a(d/dt)(1/a)$.
Equation (\ref{eq_compare}) suggests that rotational diffusion occurs on a 
time scale that is of order $\m12/\mf$ times the hardening time,
or that the binary will rotate by an angle of order $\sqrt{\mf/\m12}$
in the time it takes to harden significantly.
This result will be assessed more quantitatively in \S 5.

\clearpage

\section{Numerical Computation of the Diffusion Coefficient}

The same set of scattering experiments described in Paper I were
used here to evaluate 
$\langle\Delta \ell_x^2\rangle$ and $\langle\Delta \ell_y^2\rangle$.
As described more fully in that paper, 
field stars were treated as massless particles moving in the 
potential of the binary whose center of mass and orbital parameters
remained fixed.
Orbits were integrated using the routine D0P853 of \citet{hnw91}, 
an 8(6)th order embedded 
Runge-Kutta integrator; each field star was assumed to begin at 
a position $(x,y,z)=(\infty,p,0)$ and was advanced from $r=\infty$ to 
$r=50a$ along a Keplerian orbit about a point mass $\m12$.
The integrations were terminated when the star had moved a distance 
from the binary that was at least $100$ times its initial distance with 
positive energy.
The binary's orbit was assumed to be circular;
the orientation of the binary's orbital plane with respect to the initial
velocity vector of the field star,
and the initial phase of the binary, were chosen randomly for each integration.
Unless otherwise indicated, distances and velocities are given below in 
program units of $a$ and $(G\m12/a)^{1/2}$ respectively.
The orbit integrations, and the expressions given above 
relating changes in the field-star's angular momentum to the orientation
of the binary, were checked by carrying out a limited set of experiments
using a fully general, three-body integrator and various values for
$\mf/\m12$.
The expressions derived above for $\delta\vartheta$ etc.
were found to be accurately reproduced
in the limit of small $\mf/\m12$.

Consider first the rotational diffusion coefficient corresponding
to interactions with stars of a single velocity $V$ at infinity.
Multiplying $(\delta\vartheta)^2$ in a single encounter by $2\pi p\nf V$,
with $p$ the impact parameter and $\nf$ the density of field stars,
and integrating over $p$ gives
\begin{mathletters}
\begin{eqnarray}
\langle\delta\vartheta^2(V,p_{max})\rangle &=& 2\pi\nf V\int_0^{p_{max}} 
dp\ p\ \overline{\delta\vartheta^2}  \\
&=& 2\pi\nf V \int_0^{p_{max}} dp\ p\ 
\left({\overline{\delta \ell_{b,x}^2} + \overline{\delta \ell_{b,y}^2}\over \ell_b^2}
\right) \\
&=& {32\pi\nf V\over G\m12 a}\left({\mf\over\m12}\right)^2 
\int_0^{p_{max}} dp\ p\ 
\left(\overline{\delta \ell_x^2} + \overline{\delta \ell_y^2}\right)
\end{eqnarray}
\end{mathletters}
(cf. equation \ref{eq_diff2}), where
$\overline{\delta \ell_x^2}, \overline{\delta \ell_y^2}$ 
are defined as mean square
changes at fixed $(V,p_{max})$,  
averaged over many encounters in which the orientation of 
${\bf V}$ with respect to the binary's spin axis is taken to be random.
Figure 1 shows the distribution of field-star velocity changes as
a function of $P$ for $V=0.5 V_{\rm bin}$.

\footnotesize
\begin{figure}
\epsscale{0.8}
\plotone{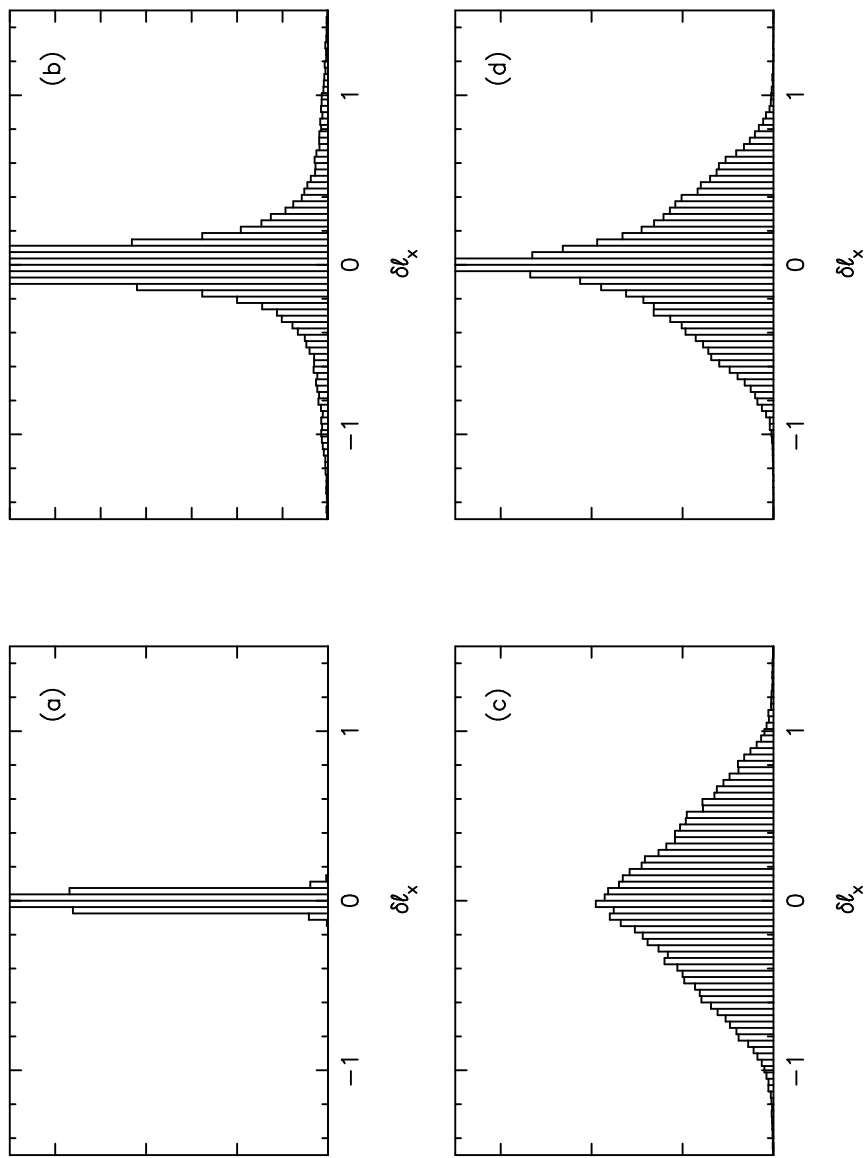}
\caption{
Distribution of field star angular momentum changes for scattering experiments
with $V/V_{bin}=0.5$.
Units are $G=\m12=a=1$.
Each plot corresponds to $5\times10^4$ scattering experiments within some 
range of impact parameters $[p_1,p_2]$ in units of $a$.
(a) $[4,6]$; (b) $[2,4]$; (c) $[1,2]$; (d) $[0.6,1]$.
}
\end{figure}
\normalsize

We define a dimensionless, velocity-dependent
diffusion coefficient $D_R$ as
\begin{mathletters}
\begin{eqnarray}
\langle\delta\vartheta^2(V,p_{max})\rangle 
&=& \pi a^2\nf V_{\rm bin} \left({\mf\over\m12}\right)^2D_R(V,p_{max}),\\
D_R(V,p_{max}) &\equiv& 16 \left({V\over V_{bin}}\right) 
\int_0^{p_{max}/a} d\left({p\over a}\right)^2 
{\left(\overline{\delta \ell_x^2} + \overline{\delta \ell_y^2}\right)\over a^2 V_{bin}^2}
\label{eq_dr}
\end{eqnarray}
\end{mathletters}
with $V_{\rm bin}=\sqrt{G\m12/a}$.
Figure 2 plots $D_R$ as a function of $V$ and $p_{max}$.
The most effective encounters are those with small $V$ and $p_{max}$;
encounters with impact parameters $p\gap \sqrt{2G\m12a}/V$ have a 
closest-approach distance to the binary of $r_p\gap a$ and produce
almost no change in the binary's orientation.
Hence $D_R$ reaches a maximum value at  $p\approx \sqrt{2G\m12 a}/V$
before levelling off at larger $p$.

\begin{figure}
\epsscale{0.8}
\plotone{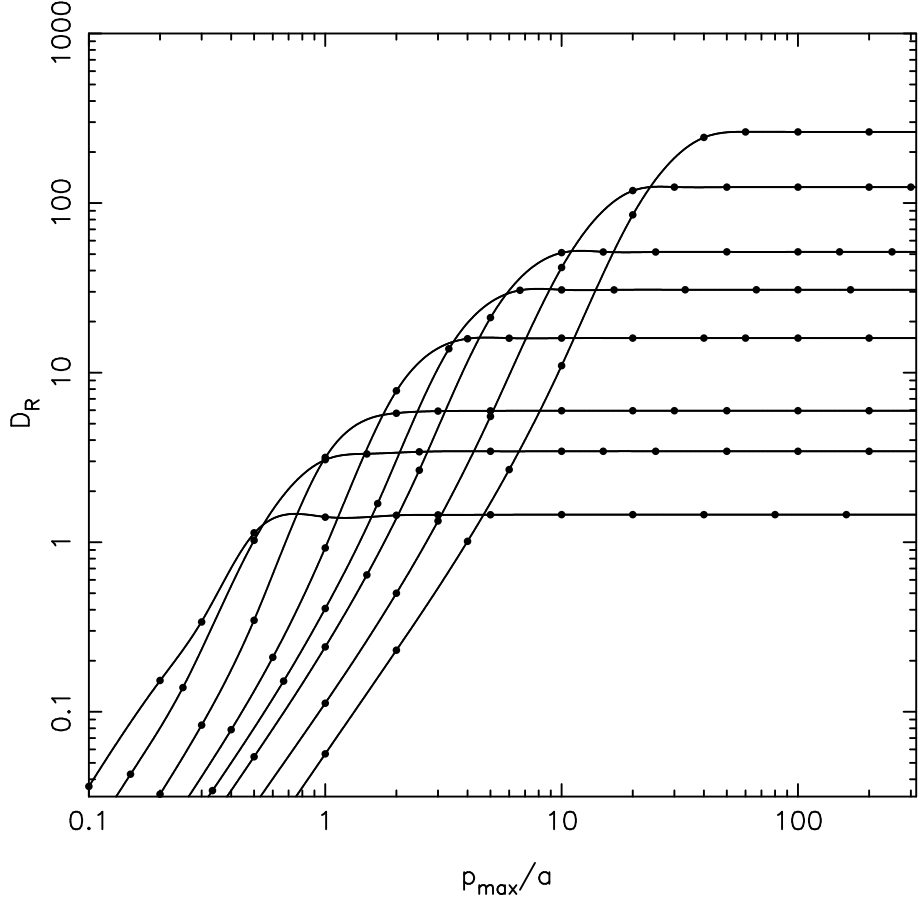}
\caption{
Velocity-dependent rotational diffusion coefficient $D_R(V,p_{max})$ 
defined in equation (\ref{eq_dr}).
$p_{max}$ is the maximum impact parameter for encounters and $a$ is the 
binary semi-major axis.
Different curves correspond to relative velocities at infinity of
$V/V_{bin} = 0.05,0.1,0.2,0.3,0.5,1,2,10$; 
the lowest (highest) $V$ produces the highest (lowest) 
$D_R$ at large $p_{max}$.
Points are averages computed from the numerical integrations; 
curves are spline fits.
}
\end{figure}

Multiplying $\langle\delta\vartheta^2\rangle$ by 
$f_f(\vf)$, the distribution of field-star
velocities, and integrating ${\bf dv}_f$ gives $\dv$.
We assume a Maxwellian distribution of field star velocities,
\beq
\ff(v_f) = {1\over (2\pi\sf^2)^{3/2}} e^{-v_f^2/2\sigma_f^2},
\label{max}
\eeq
so that
\beq
\dv = {4\pi\over\left(2\pi\sigma^2\right)^{3/2}}\int_0^{\infty} dV\ V^2 e^{-V^2/2\sf^2}\langle\delta\vartheta^2\rangle .
\label{eq_defl1}
\eeq
This can be written in terms of a dimensionless coefficient $L$,
\begin{mathletters}
\begin{eqnarray}
\dv&=& {\mf\over\m12}{G\rho_f a\over\sf}L, \label{eq_defdv} \\
\label{eq_defla}
L &\equiv& \sqrt{2\pi} S^{-1} \int_0^{\infty} dz\ z^2 e^{-z^2/2} D_R(zV_{\rm bin}/S,R\pf).
\label{eq_deflb}
\end{eqnarray}
\end{mathletters}
$L$ is defined in analogy with the dimensionless coefficients 
$(H,J,K)$ that describe respectively the rates of change 
of the binary's energy, the rate of mass ejection by the binary, and the
binary's eccentricity growth rate (cf. \citet{qui96}).
Note that $L$ is a function of two parameters, $L=L(R,S)$, where
\beq
R\equiv {p_{max}\over\pf}={p_{max}\sf^2\over G\m12},\ \ \ \ S\equiv{V_{bin}\over\sf}.
\eeq
$R$ is the maximum impact parameter in units of the radius of gravitational
influence of the binary and $S$ is the dimensionless hardness of the binary.
Figure 3 plots $L(R,S)$.
At a given hardness $S$, $L$ reaches its limiting value 
$L_{\infty}\equiv L(\infty,S)$ by $R\approx 3/S$.
For all but the softest binaries, this limiting value is reached before
$R=1$, which is roughly the value of $R$ we expect for a massive binary at the center of a steeply-falling density cusp (Paper I).
Hence we ignore the dependence of $L$ on $p_{max}$ in what follows and
write $L(S)=L({\infty},S)$.
$L(S)$ varies from $\sim 25$ for the softest binaries ($S\approx 1$)
to $\sim 60$ for $S\approx 10$.

\begin{figure}
\epsscale{0.9}
\plotone{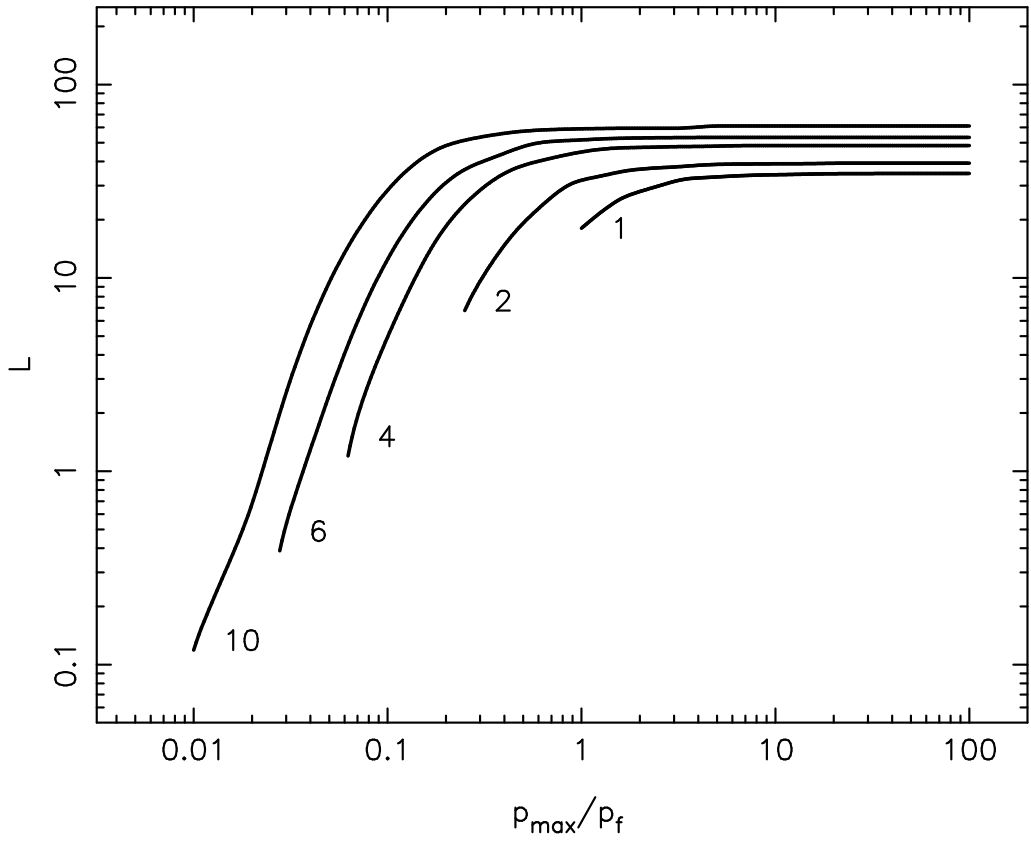}
\caption{
Rotational diffusion coefficient $L$ (equation 20) for a 
circular-orbit binary.
Curves are marked by  $V_{\rm bin}/\sf=1,2,4,6,10$.
}
\end{figure}

\section{Solutions of the Evolution Equation}
Equation (\ref{eq_evol1}) describes the evolution of $F(\theta,t)$,
the probability distribution for the binary's orientation, given
the rotational diffusion coefficient $\dv$.
The latter (equation 20) is a function of the binary separation 
$a$, directly via the factor $G\rf a/\sf$,
 and indirectly via the weak dependence of $L$ on binary hardness (Figure 3).
Furthermore the timescale for the binary to reorient itself is shorter than 
the hardening time $\left|(1/a)(da/dt)\right|^{-1}$ 
(cf. equation \ref{eq_compare}), 
hence the change in $a$ can not be neglected when solving for the 
evolution of the orientation.

We nevertheless begin by considering an idealized model in which 
$\dv$, $a$, and the parameters $(\rf,\sf)$ that describe the background
are assumed constant in time.
The evolution equation (\ref{eq_evol1}) becomes
\beq
{\partial F\over\partial\tau} = {1\over 2} {\partial\over\partial\mu} \left[\left(1-\mu^2\right){\partial F\over\partial\mu}\right],
\eeq
where $\mu\equiv\cos\theta$, $\tau\equiv t/t_0$, and
\beq
t_0 \equiv 2\left({\m12\over\mf}\right){\sf\over G\rf a L}.
\eeq
Consider the evolution of $\overline{\mu}(t)=2\pi\int_{-1}^{1}F(\mu',t)\mu' d\mu'$, the expectation value of $\mu$.
We have
\begin{mathletters}
\begin{eqnarray}
{d\overline{\mu}\over d\tau} &=& 2\pi\int_{-1}^1
{\partial F\over\partial\tau} \mu' d\mu' \\
&=& \pi\int_{-1}^1 \mu'{\partial\over\partial\mu'}\left[\left(1-\mu'^2\right){\partial F\over\partial\mu'}\right] \\
&=&-2\pi\int_{-1}^1\mu' F d\mu' = -\overline{\mu}
\end{eqnarray}
\end{mathletters}
or
\beq
\overline{\mu}(\tau) = \overline{\mu}_0 e^{-\tau}.
\eeq
Hence $t_0$ is the time constant for relaxation to a uniform distribution 
of orientations, $\overline{\mu}=0$, 
given a fixed binary separation and unchanging stellar background.
It may be expressed in physical units as
\beq
t_0 = 6.1\times 10^{9} {\rm yr} \left({\m12\over 10^6 \mf}\right)\left({L\over 50}\right)^{-1} 
\left({\rf\over 10^6\msun {\rm Mpc}^{-3}}\right)^{-1} 
\left({\sf\over 200\ {\rm km}\ {\rm s}^{-1}}\right)   
\left({a\over 1\ {\rm pc}}\right)^{-1}.
\eeq

Next consider the more interesting case where $a$, as well as the
parameters $\rf$ and $\sf$ that describe the field stars, 
may be changing with time.
This time dependence is not easily specified but we can make progress
by changing evolution variables from $t$ to 
$x\equiv\log(a/a_0)$.
First combining equations (\ref{eq_evol1}) and (\ref{eq_defdv}), we find
\beq
{\partial F\over\partial t} = {L\over 4}{\mf\over\m12}{G\rf a\over\sf} 
{\partial\over\partial\mu} \left[\left(1-\mu^2\right){\partial F\over\partial\mu}\right].
\eeq
Now changing variables,
\beq
{\partial F\over\partial t}={\partial F\over\partial x}{dx\over dt} = 
{\partial F\over\partial x} {1\over a}{da\over dt} = 
-{\partial F\over\partial x} {G\rf Ha\over\sf}
\eeq
(cf. equation (\ref{eq_defH}))
and so
\beq
{\partial F\over\partial x} = -{1\over 4} {L\over H} {\mf\over\m12} {\partial\over\partial\mu}\left[\left(1-\mu^2\right){\partial F\over\partial\mu}\right].
\eeq
This equation gives the evolution of the binary's orientation in terms of 
changes in its semi-major axis $a$, 
with no explicit dependence on the parameters that describe the 
stellar background.
Ignoring the weak dependence of $H$ and $L$ on $a$, 
appropriate for a hard binary, 
we find as before
\beq
\overline{\mu}(a) = 
\overline{\mu}_0\left({a\over a_0}\right)^{{1\over 2}{L\over H}{\mf\over\m12}}.
\eeq

The exponent in this expression is of order $\mf/\m12\ll1$. 
Hence we can write
\beq
\overline{\mu}(a)\approx 1 + {1\over 2}{L\over H}{\mf\over\m12}\log\left({a\over a_0}\right)
\eeq
where $\overline{\mu}_0$ has been set to unity, corresponding to an initial orientation parallel to the $\theta=0$ axis.

For the case of binary supermassive black holes, 
if we define $a_0$ as the separation when the binary first forms a bound
pair, we expect gravitational radiation coalescence to occur when 
$a/a_0\approx 10^{-2}$ \citep{mer00}.
Since $L/2H\approx 2$ for a hard binary, we find for the expectation
value of $\mu$ at coalescence:
\beq
\overline{\mu}\approx 1 - 10{\mf\over\m12}.
\label{eq_approx}
\eeq
Writing $\delta\theta \equiv \sqrt{2(1-\overline{\mu})}$, 
the rms change in the angle defined by the binary's spin axis, 
this becomes
\beq
\delta\theta\approx \sqrt{20\mf\over\m12},\ \ \ \ \ \ \mf\ll\m12.
\eeq
For $\mf/\m12=10^{-6}$, e.g. a binary of mass $10^6\msun$ 
surrounded by $1\msun$ stars, this predicts $\delta\theta\approx 0.25^{\circ}$.
For a binary of so-called ``intermediate-mass'' black holes with
$\m12\approx 10^3\msun$, $\delta\theta\approx 8^{\circ}$.

Milosavljevic \& Merritt (2001) give
Information about the rotational Brownian motion of a massive binary
in a set of $N$-body simulations.
They show (in their Figure 11) the evolution of 
the orientation of an equal-mass, circular-orbit binary in three
simulations with $\mf/\m12=\{164,328,655\}$.
The simulations all follow the evolution of the binary until 
$a/a_0\approx 10^{-1}$; hence equation (\ref{eq_approx}) predicts 
$\delta\theta\approx\{14^{\circ},10^{\circ},7^{\circ} \}$.
These values are in excellent agreement with the angular deflections shown 
in Figure 11 of Milosavljevic \& Merritt (2001).

Under what circumstances could the roeorientation angle of a
supermassive black hole binary be much greater than the modest values 
predicted here?
Since $\delta\theta\propto \mf^{1/2}$, one possibility is for 
the perturbing objects to be much more massive than $1\msun$; 
for instance, they could be supergiant stars
($\mf\approx 10^2\msun$), star clusters ($\mf\approx 10^3\msun$), 
or giant molecular clouds ($\mf\approx 10^4\msun$).
Such physically large objects would presumably produce most
of their perturbing torques when the binary separation was greater
than or comparable to their own size.

An even more extreme possibility is that the perturbing objects
are themselves supermassive black holes.
Consider for example a binary supermassive black hole
at the center of a giant elliptical galaxy.
The stellar densities at the centers of such galaxies are low and
one might expect the coalescence of the binary to stall due to a shortage
of interacting stars (e.g. \citet{mim01}).
In this case, further decay of the binary might await the infall of
a third and subsequent black holes from later galactic mergers
(e.g. \citet{val96}).
Mass ratios $\mf/\m12$ would then be of order $10^{-2}-10^0$ and signficant
reorientation of the binary could occur before coalescence.

Orientations of radio jets in Seyfert galaxies are known to be almost 
random with respect to the plane of the stellar disk (e.g. \citet{kin00}).
If jets are launched parallel to the spin axes of rotating black holes,
this implies that the spins of nuclear black holes bear almost
no relation to disk spins.
While the modest reorientation rates derived here are probably
insufficient to explain this phenomenon, a related mechanism may
suffice.
The spin orientation of a supermassive black hole that forms via 
{\it repeated} mergers with smaller black holes would undergo a random walk 
due to the random orbital inclinations of the infalling black holes.
If nuclear black holes in Seyfert galaxies grew through multiple
mergers before the formation of the stellar disks, 
one might expect their spin axes to be uncorrelated with those
of the stars and gas.

\acknowledgments
This work was supported by NSF grants AST 96-17088 and 00-71099 and
by NASA grants NAG5-6037 and NAG5-9046.

\end{document}